\documentclass[twocolumn]{aastex631}

\usepackage{CJKutf8} 
\usepackage{amsmath}
\usepackage{comment}
\usepackage{physics}
\usepackage{textcomp}
\usepackage{multirow}

\newcommand\Msol{M$_{\odot}$}

\newcommand{\hi}{H\,{\sc i}}

\usepackage{soul}
\usepackage{threeparttable}

\shortauthors{Khim et al.}
\shorttitle{Disco Ball}

\begin{document}

\title{A Closer Look at an Unusual Ultra-Diffuse Galaxy\footnote{The data presented herein were obtained at the W. M. Keck Observatory, which is operated as a scientific partnership among the California Institute of Technology, the University of California and the National Aeronautics and Space Administration. The Observatory was made possible by the generous financial support of the W. M. Keck Foundation.}}  
\correspondingauthor{Donghyeon J. Khim}
\email{galaxydiver@arizona.edu}

\author[0000-0002-7013-4392]{Donghyeon J. Khim} 

\affiliation{Steward Observatory and Department of Astronomy, University of Arizona, 933 N. Cherry Avenue, Tucson, AZ 85721, USA} 

\author[0000-0002-5177-727X]{Dennis Zaritsky}

\affiliation{Steward Observatory and Department of Astronomy, University of Arizona, 933 N. Cherry Avenue, Tucson, AZ 85721, USA} 

\author[0000-0001-8568-8729]{Loraine Sandoval Ascencio} 

\affiliation{Department of Physics \& Astronomy, University of California, Irvine, 4129 Reines Hall, Irvine, CA 92697, USA} 

\author[0000-0003-1371-6019]{M. C. Cooper}

\affiliation{Department of Physics \& Astronomy, University of California, Irvine, 4129 Reines Hall, Irvine, CA 92697, USA} 

\author[0000-0001-7618-8212]{Richard Donnerstein} 

\affiliation{Steward Observatory and Department of Astronomy, University of Arizona, 933 N. Cherry Avenue, Tucson, AZ 85721, USA}

\thanks{Corresponding author: \href{mailto:galaxydiver@arizoan.edu}{galaxydiver@arizona.edu}}

\thanks{$^*$ The data presented herein were obtained at the W. M. Keck Observatory, which is operated as a scientific partnership among the California Institute of Technology, the University of California, and the National Aeronautics and Space Administration. The Observatory was made possible by the generous financial support of the W. M. Keck Foundation.}

\begin{abstract}
We present a spectroscopic study of the ``Disco Ball'' (SMDG0038365-064207), a rotationally-supported, red-sequence, ultra-diffuse galaxy (UDG) with a nuclear star cluster (NSC), multiple stellar clusters, and active star-forming regions using data obtained with KCWI on the Keck II Telescope. We calculate that the galaxy hosts $34\pm11$ ``globular" clusters.
Kinematic measurements confirm rotation with a peak rotational velocity of at least 53km s$^{-1}$ and a dynamical mass within $r_{\rm e}$ of at least $10^{9.3 \pm 0.2}$\Msol. 
Our dynamical estimates of the halo mass are consistent with that obtained using the number of globular clusters and together suggest $M_{\rm h}=10^{11.1\pm0.2}$\Msol.
The NSC may exhibit signatures of weak AGN activity. Our findings challenge two common assumptions: (1) clusters in some UDGs may be younger than generally assumed, and thus more luminous than standard globular clusters (GCs), affecting GC counts and the derived GC luminosity function in these UDGs, and (2) quiescent UDGs can be rotationally supported, making kinematic measurements viewing-angle dependent in such cases. The Disco Ball, while unremarkable in mass, size, projected structural properties, or color, reveals surprising complexity, highlighting the need for detailed studies of more UDGs.

\end{abstract}


\section{Introduction}
\label{sec:intro}

Ultra-diffuse galaxies (UDGs), as defined by \cite{2015vanDokkum}, 
are large (effective radius, $r_{\rm e} > 1.5$ kpc), low surface brightness (central surface brightness in the $g-$band, $\mu_{0,g} \ge 24$ mag arcsec$^{-2}$) galaxies. Their extreme properties challenge our understanding of galaxy formation and evolution. Because any challenge is an opportunity to develop a more complete understanding, they are a current focus of study and discussion \citep[e.g.,][]{Yozin_2015,Amorisco2016,DiCintio2017,Chan2018,Jiang2019,Sales_2020,Wright2021,Benavides_2023,Fielder2024}.
We present an unusual UDG, SMDG0038365-064207, which we refer to as the ``Disco Ball" (the rationale for this name will be become apparent). This galaxy stands out even among this diverse class of galaxies, offering new insights into UDG evolution.

UDGs found in galaxy clusters are typically red and quenched \citep[e.g.,][]{2015vanDokkum,koda}, while field UDGs are generally blue and star-forming \citep[e.g.,][]{prole, greco, leisman,kadowaki21}. Larger samples \citep{smudges5,barbosa} hint at more complex, perhaps episodic, behavior and \cite{loraine} find examples of quenched, starforming, and post-starburst among a set of UDGs outside of galaxy clusters. As we will show, the Disco Ball is a hybrid case, with numerous small knots of star formation but otherwise consisting of a diffuse, quiescent stellar population. It lies in an unremarkable environment, with the nearest spectroscopically-confirmed neighbor projected $\sim$1.7Mpc away. 
The Disco Ball demonstrates that a low level of star formation can proceed in UDGs in the absence of any clear environmental drivers. 

Some UDGs possess surprisingly rich populations of globular clusters (GCs) given their relatively low stellar masses \citep{Beasley_2016,2017vanDokkum,Lim2018,Toloba_2018,Somalwar_2020,Gannon2022}. For galaxies in general, the number of GCs is proportional to the galaxy's total mass \citep{blakeslee, Spitler2009, Harris2017}. As such, the large numbers of GCs found in at least some UDGs have been used to infer more massive dark matter halos than expected for galaxies of corresponding stellar mass. However, there is evidence of unusual behavior in GC-richness among UDGs, perhaps related to environment or dark matter fraction \citep{forbes+18, Somalwar_2020,Gannon2022}, and for overly luminous GCs in some \citep{Shen_2021,janssens,tang,li}. The nature of GC formation and evolution in UDGs is therefore somewhat unclear. As we will show, the Disco Ball has a varied population of several tens of luminous clumps that are consistent with being young and intermediate age stellar clusters. As such, this galaxy highlights how the UDG cluster population can evolve and how, at least in some UDGs, GC-like objects can continue to form up to the current time even though the overall level of star formation is low and the bulk of the stellar population in the galaxy is diffuse.

\begin{figure*}
	\includegraphics[width=\linewidth]{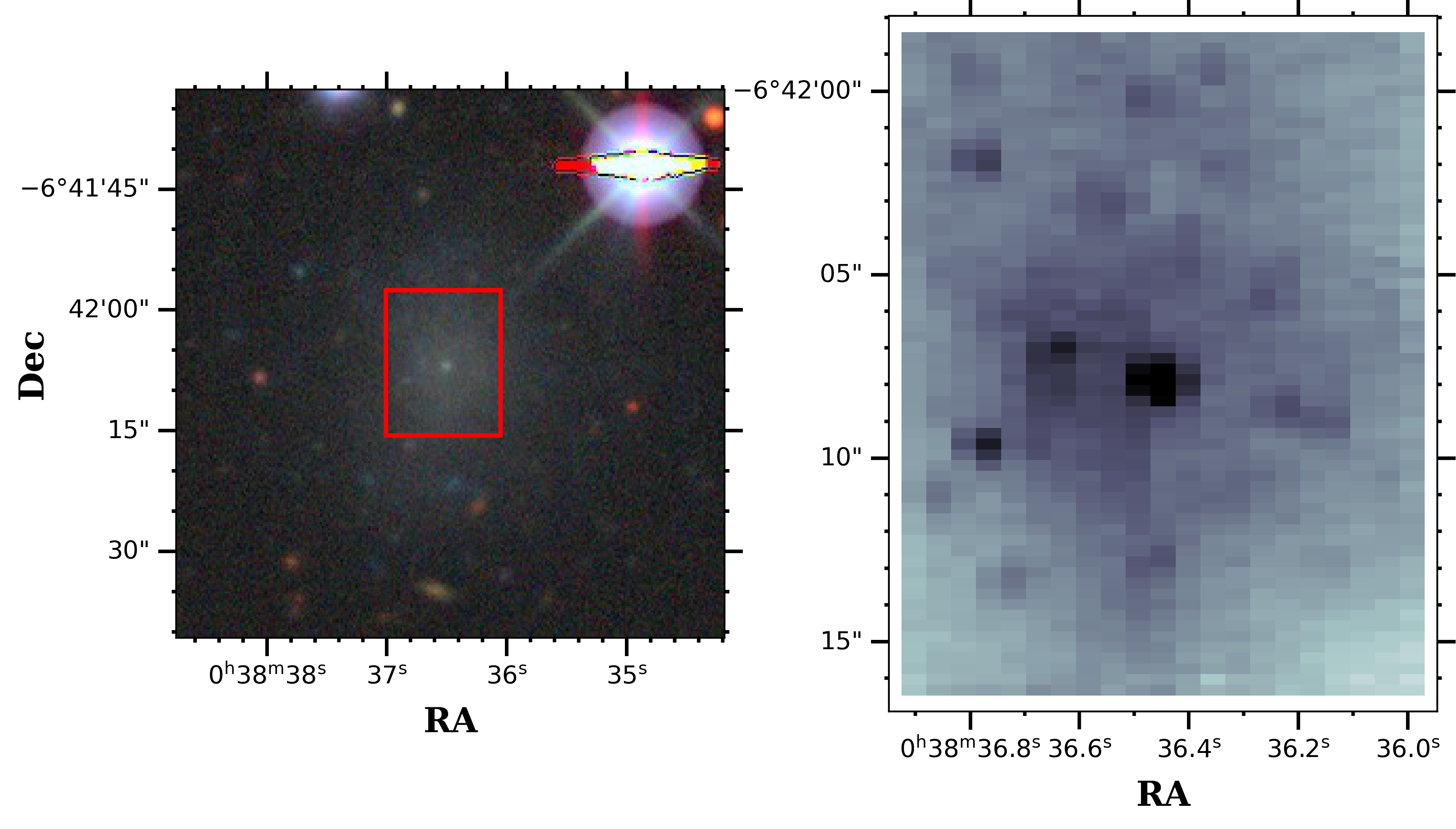}
    \caption{\textit{Left}: Legacy Surveys image of our target galaxy (``Disco Ball" or SMDG0038365-064207). The image is drawn from the Legacy Surveys online viewer (\url{ https://www.legacysurvey.org/viewer}). The field of view of the image is 68.1 arcsec across (covering approximately 2$r_{\rm e}$ from the center). The red box indicates the field of view of our KCWI data. We only have spectroscopy for the galaxy's inner region (within the red box), though there are additional stellar clumps outside this area. \textit{Right}: KCWI images of our target galaxy. For each ``spaxel'' (See \S\ref{sec:reduction}), we stack pixel values across the entire wavelength to create this image. The field of view of the image is 16.5 $\times$ 20.4 arcsec, covering out to about 0.6$r_{\rm e}$ from the center. }
    \label{fig:image_legacy}
\end{figure*}

\begin{figure}
	\includegraphics[width=\columnwidth]{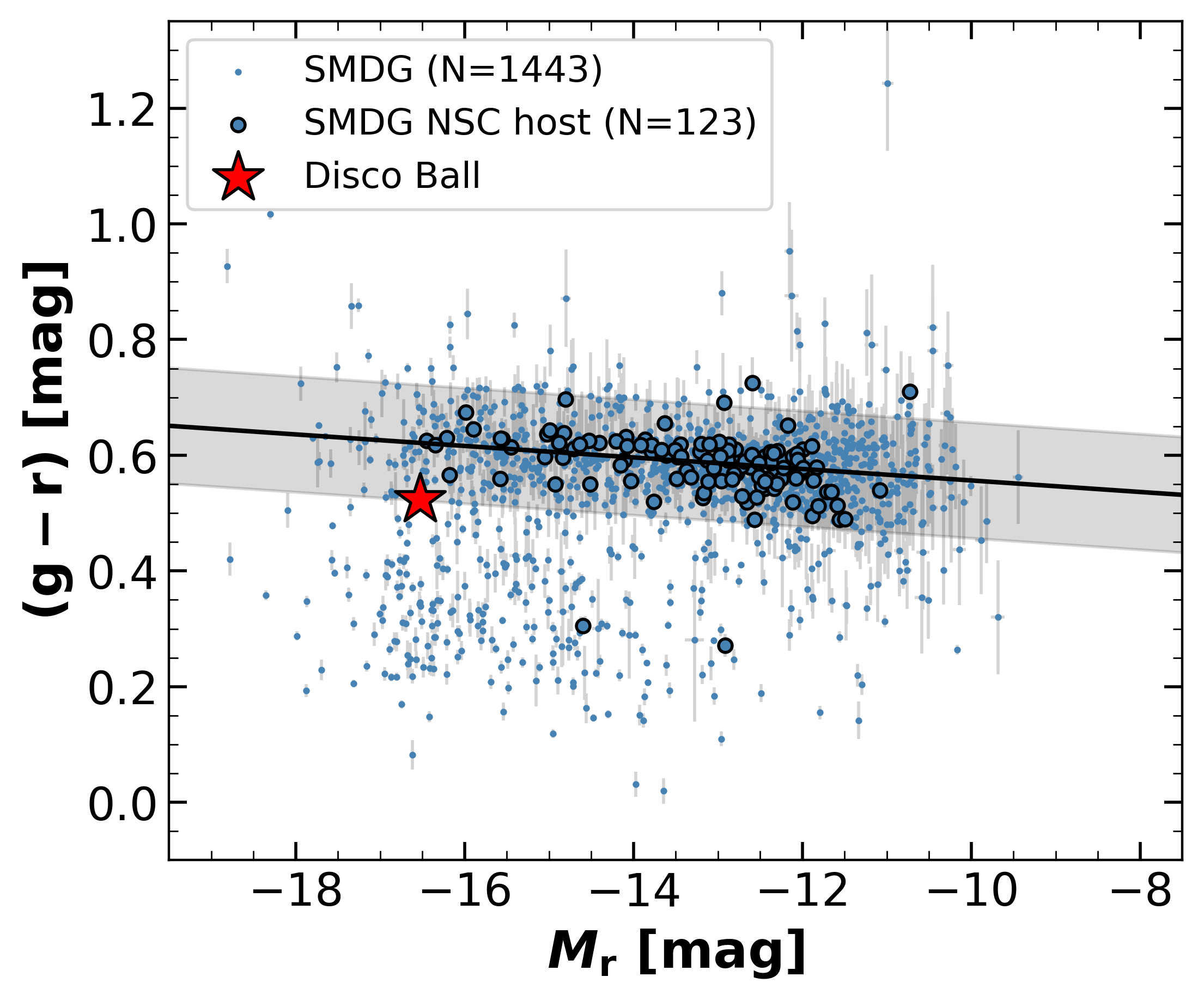}
    \caption{Color-magnitude relation for SMUDGes with distance estimates \citep{khim+24_nsc}. The black line represents the fitted red sequence. Galaxies whose color difference from the red sequence is less than 0.1 mag are considered to be red sequence galaxies (shaded region). Galaxies with an identified NSC are represented by circles and are almost exclusively found along the red sequence.  Our target, the Disco Ball, is represented by a red star and is located at the blue edge of the red sequence.}
    \label{fig:cmd}
\end{figure}

A modest fraction of UDGs ($\gtrsim 10$\%) host nuclear star clusters (NSCs) \citep{lambert,khim+24_nsc}. NSCs are extremely compact (2–40 pc) and typically much brighter than ordinary GCs \citep{Cote2006}. NSCs are found at somewhat higher rates in slightly more massive galaxies than in the SMUDGes UDGs \citep[in the stellar mass range of $10^{8}$ -- $10^{10}$\Msol;][]{denBrok2014, Ordenes2018, neumayer} and are thought to form either through the migration and merging of star clusters \citep[e.g.,][]{tremaine1975, lotz, CDMB2009, gnedin} or the nuclear accretion of gas \citep[e.g.,][]{Bailey1980, mihos94, bekki01, Seth2006, Walcher2006}. The potential connection between UDG evolution and an NSC is unexplored.  
Broadly, the most massive NSCs are postulated to be likely hosts of intermediate-mass black holes \citep{pz,antonini} and therefore may at times be active galactic nuclei and affect the host galaxy via feedback. As we will show, there is tentative evidence of weak activity in the nucleus in the Disco Ball.

The galaxy that is our current focus was originally identified as a candidate UDG in 
the SMUDGes survey \citep{smudges, smudges5}. SMUDGes cataloged thousands of UDG candidates, offering an opportunity to study statistically significant samples of subclasses of these enigmatic galaxies. 
\cite{loraine} present spectroscopic follow-up using the Keck Cosmic Web Imager (KCWI) on Keck II \citep{kcwi_martin, kcwi} of 44 SMUDGes galaxies located outside of the cluster environment, focusing on the nature of the recent global star formation behavior of UDGs. 
From among these 44 galaxies, we highlight the Disco Ball as a particularly interesting case (Figure \ref{fig:image_legacy}). 
Aside from a central point source (NSC), it appears to be a rather typical diffuse UDG candidate in the Legacy Surveys imaging \citep{dey}. In hindsight,
luminous clumps are hinted at in the Legacy Survey image (left panel), but they are not nearly as evident as in the KCWI image (right panel). Moreover, with the understanding gained from the KCWI image, it appears from the Legacy Survey image that there are additional clumps that are likely associated with this galaxy outside the KCWI field of view. Due to the number and distribution of clumps, we refer to this galaxy as the ``Disco Ball".

With the KCWI measured redshift \citep[$z = 0.01067$;][]{loraine}, the basic properties of this galaxy are now known (Table \ref{table:properties}). Using photometric measurements from  \cite{smudges5} and \cite{khim+24_nsc}, who measure $m_g = 17.37$, $g-r = 0.51$, and $r_{\rm e} = 15$ arcsec, 
we place the galaxy on the color-magnitude diagram (Figure \ref{fig:cmd}), determine that it has an absolute magnitude of $M_r = -16.7$ mag, and confirm it as a UDG with $r_{\rm e} = 3.3\pm0.1$ kpc for our adopted standard $\Lambda$CDM cosmology \citep{wmap9}. It is one of the largest and brightest of the SMUDGes UDGs. The galaxy is classified as a `red-sequence' galaxy based on its position in the color-magnitude diagram, but it lies at the blue edge of that population. Finally,
\cite{khim+24_nsc} report that the NSC has an $r$-band magnitude of 22.75 mag, corresponding to an absolute magnitude of $M_r = -10.59$ mag. Approximately 0.4\% of the stellar mass of the host galaxy lies in the NSC. This fraction is not unusual for such systems \citep{khim+24_nsc}. We list all of these properties and a few more, such as S\'ersic index ($n$), axis ratio ($b/a$), and major axis position angle position angle (PA) in Table \ref{table:properties}. 

\begin{deluxetable}{lrr}
\tablewidth{0pt}
\tablecaption{Properties of the Disco Ball$^a$
\label{table:properties}}
\tablehead{
Property&Value&Units\\
}
\startdata 
RA & 00:38:36.5 &(J2000.0)\\
DEC & $-$06:42:07 & (J2000.0) \\
Redshift & 0.01067 & unitless \\
$D_L$ & 46.5 & Mpc\\
$D_A$ & 45.6 & Mpc \\
$\mu_{0,g}$ & 24.23$^{+0.03}_{-0.15}$ & mag arscec$^{-2}$\\
$\mu_{0,r}$ & 23.64$^{+0.04}_{-0.13}$ & mag arscec$^{-2}$\\
$m_g$ & 17.37$^{+0.06}_{-0.18}$ & mag \\
$m_r$ & 16.80$^{+0.06}_{-0.18}$ & mag\\
$r_{\rm e}$ & 15.0$^{+0.3}_{-0.3}$ & arcsec\\
$r_{\rm e}$ & 3.3$^{+0.1}_{-0.1}$ & kpc\\
$M_g$ & $-$16.12$^{+0.06}_{-0.18}$ &  mag\\
$M_r$ & $-$16.63$^{+0.06}_{-0.18}$ &  mag\\
b/a & 0.75$\pm0.02$ & unitless\\
PA & $-8 \pm 2$ & degrees (N to E)\\
$n$ & 0.91$^{+0.02}_{-0.02}$ & unitless \\
$A_g$ & 0.18 & mag\\
$A_r$ & 0.12 & mag\\
NSC $m_r$ & 22.75$^{+0.03}_{-0.03}$ & mag\\
NSC $M_r$ & $-$10.59$^{+0.03}_{-0.03}$ & mag 
\enddata
\tablenote{Values adopted from  \cite{smudges5},   \cite{loraine}, and \cite{khim+24_nsc} or calculated using values from those references. Absolute magnitudes include Galactic extinction corrections using the listed $A_g$ and $A_r$ values.}
\end{deluxetable}

We utilize the spectroscopic nature of the data, as well as the superior resolution and sensitivity of KCWI relative to the Legacy Survey image (Figure \ref{fig:image_legacy}), to investigate the nature of these stellar clumps and to better understand the internal kinematics and stellar populations of the Disco Ball. We aim to understand the connections between UDGs, their stellar clusters, star formation behavior, and, when present, a nuclear star cluster.
This paper is structured as follows.  We outline our methodology, detailing the processes for observing and reducing the data, including the removal of residual artifacts in \S\ref{sec:method}, describe the fundamental parameterization, cluster finding, and spectral analysis, including the measurement of velocities, velocity dispersions, and dynamical masses in \S\ref{sec:analysis},
present our results
in \S\ref{sec:results}, discuss some of the implications of our measurements in \S\ref{sec:discussion}, and summarize our key conclusions in \S\ref{sec:summary}. Throughout this work, we adopt a standard WMAP9 cosmology \citep{wmap9}. Our results are robust to the current range of uncertainties in cosmological parameters. Magnitudes are given in the AB system \citep{oke1,oke2}.

\section{Methodology}
\label{sec:method}

\subsection{Data Acquisition and Reduction}
\label{sec:reduction}

We observed the Disco Ball, as part of a KCWI spectroscopic follow-up of a set of field SMUDGes candidate UDGs \citep{loraine}. The observations of the Disco Ball occurred during the fall of 2021, where we employed the image slicer with the BL grating, centered at 4500\AA\  and covering a spectral range from 3500 to 5500\AA. The resulting data extend to about one-half of the target's effective radius (the field of view is 16 arcsec $\times$ 20 arcsec). The exposure time was 1800 seconds. We acquire a sky frame, offset by approximately 300 arcsec with total exposure time matching the science frames to facilitate non-local sky subtraction, as well as standard calibration frames, including bias, arc, and flat frames. We reduce the data using the KCWI Data Reduction Pipeline\footnote{\url{https://kcwi-drp.readthedocs.io/en/latest/}}. Readers may refer to \cite{loraine} for a more detailed description of the data and the data reduction.

Regarding our nomenclature, we refer to a ``spaxel" as any array element in the KCWI data that contains the spectrum corresponding to a specific location within the field of view. We refer to a ``pixel" as any detector element along the spectrum contained within a spaxel.

After standard data reduction, which includes cosmic ray rejection, some pixels remain affected by cosmic rays. We apply sigma clipping, removing from further consideration any pixel values that lie $\ge$ 20$\sigma$ away from the mean.
The rejected pixels are limited to specific spaxels and wavelength ranges, as would be expected in the case of cosmic rays. Using this process, we reject 83 pixels out of 19,341,240.

\begin{figure}
	\includegraphics[width=\columnwidth]{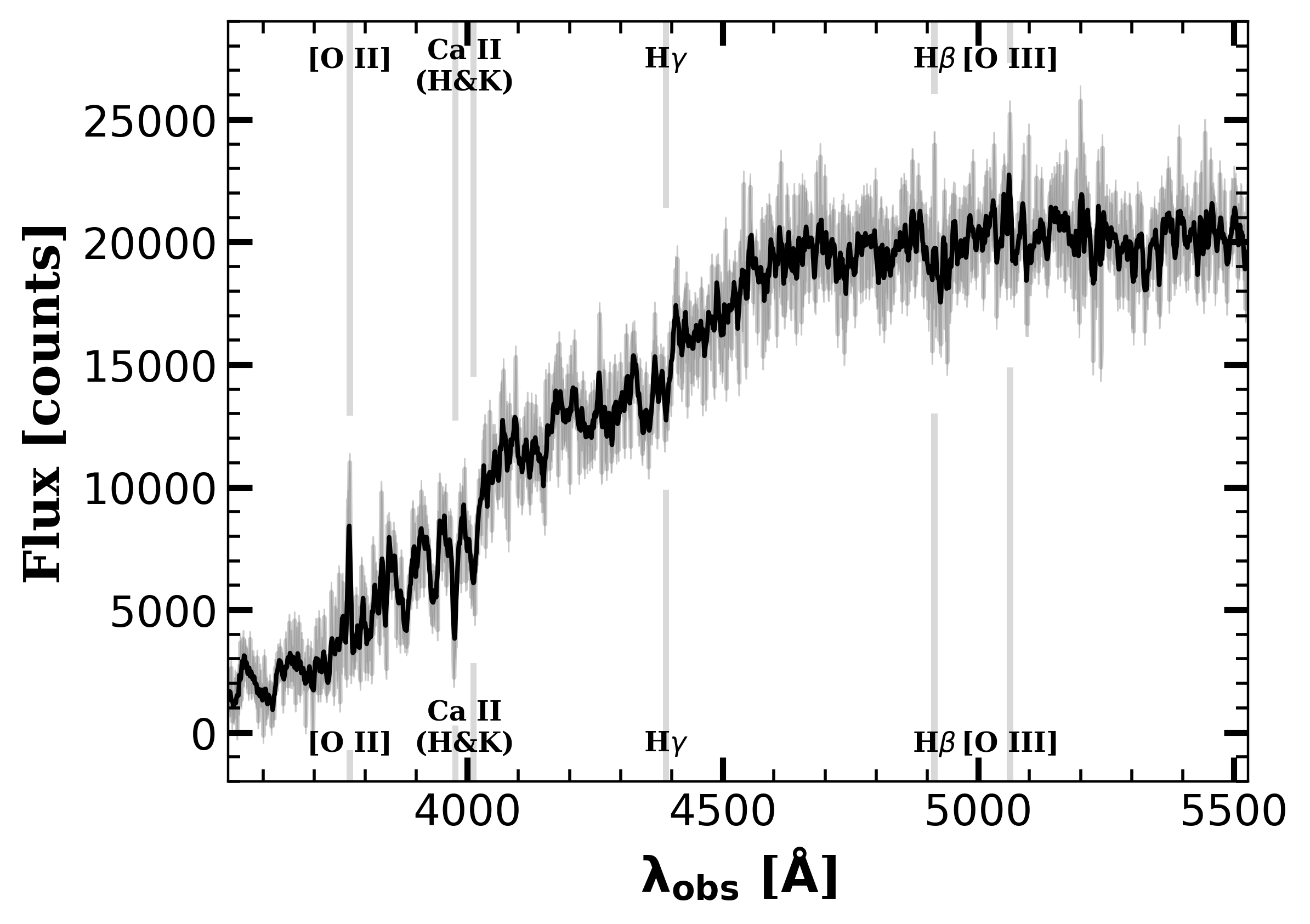}
    \caption{The stacked spectrum of all spaxels is the dark line, with the shaded area indicating the uncertainty. We also mark the location of commonly observed spectral lines for reference.}
    \label{fig:spec_tot}
\end{figure}

We present the uncalibrated spectrum using all remaining pixels across all spaxels in Figure \ref{fig:spec_tot}. Emission and absorption features are visible, but overall this is a relatively quiescent galaxy dominated by intermediate and old stellar populations. 

\subsection{The Point and Spectral Line Spread Functions}
\label{sec:quality}

Neither the spaxel nor the pixel values are fully independent. The telescope, instrument, and atmosphere spread the light from a point source over several spaxels and the spectral dispersion spreads a delta function spectral line over several pixels. We need to understand both of these effects to analyze our data.
We now quantify the point spread function (PSF) and the instrumental line broadening (ILB). 

\subsubsection{Point Spread Function}
\label{sec:psf}

The typical effective radius of star clusters, including NSCs, can be as large as 40 pc \citep{Georgiev2016, neumayer}, indicating that the clusters remain unresolved in ground-based data at the distance of the Disco Ball. Therefore, we expect all of our sources that are stellar clusters to have a surface brightness profile that is consistent with the PSF.

We measure the KCWI PSF using the NSC because it is the brightest point source in the image.
We fit a Gaussian profile to the NSC independently along the RA (Right Ascension) and Dec (Declination) directions because the KCWI plate scales differ along each direction, and measure the full-width half maximum (FWHM) from the fit. The measured FWHMs are 1.23 and 2.87 spaxels in the RA and Dec direction, respectively, corresponding to 0.84 $\pm$ 0.02 arcsec. The measured PSF is therefore consistent with a plausible effective seeing (i.e., less than 1 arcsec) for the reduced images.

\subsubsection{Instrumental Line Broadening}
\label{sec:dispersion}

To measure the ILB, 
we identify localized emission line regions in the image by constructing an [O III] image slice. The slice is an image created from the KCWI data using only a 5\AA\ spectral bandpass centered at the redshifted wavelength of the [O III]5007 line. The width of the 5\AA\ bandpass corresponds to about 300km s$^{-1}$, which is broad enough to include emission even when accounting for the galaxy's possible rotation and velocity dispersion. We identify the three brightest clumps (excluding the NSC) and measure their emission line profiles (the identification process will be described in the following sections). The average measured FWHM of the emission lines  in the spectral direction is 2.37 $\pm$ 0.05\AA\ (corresponding to $\sigma = 1.0$\AA\ or 68km s$^{-1}$, and, as expected, is much larger than the velocity dispersion of any likely stellar cluster). As such, we assume that the internal velocity dispersion within these localized regions is negligible relative to the ILB and assign the full measured width to the ILB. This measured value agrees with the expectation based on the KCWI documentation.

\section{Analysis}
\label{sec:analysis}

\subsection{Cluster Finding}
\label{sec:finding}

In this subsection, we describe our process for identifying luminous clumps in the galaxy. While we refer to these clumps as stellar clusters, it is important to note that we lack information on whether they are gravitationally bound or, prior to further analysis, even dominated by stellar continuum emission. Nevertheless, because there is no other known class of object involving a luminous concentration of stars other than dense star clusters in UDGs, we assume that our objects are structurally consistent with the clusters found elsewhere.

To summarize our process, we reconstruct the KCWI data into different ``slices'', images that either contain flux only from specific emission lines of interest ([O II], H$\beta$, or [O III]), or that exclude those emission lines to produce a complementary continuum ``slice". We then identify the clusters in the separate slices using the Source Extractor Python library \citep[SEP; see][for details]{sep}, which is based on Source Extractor \citep{bertin}, as described in more detail below.

\begin{figure}
	\includegraphics[width=\columnwidth]{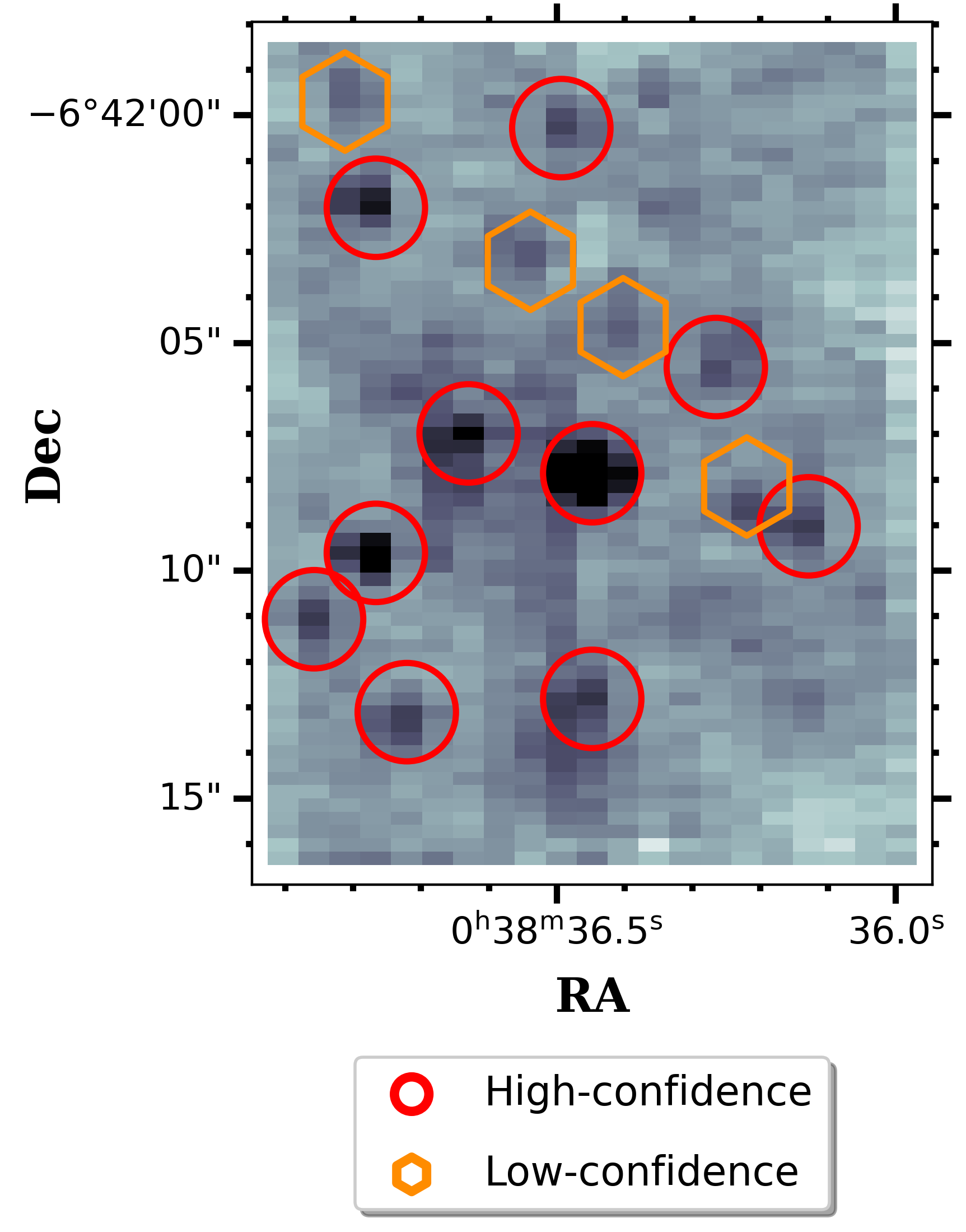}
    \caption{Cluster detection on the background-subtracted continuum slice. The red circles highlight identified objects that have at least three pixels where each pixel is $> 2\sigma$ above the background (high-confidence clusters), and the orange hexagons objects that have at least three pixels where each pixel is $> 1.5\sigma$ above the background (low-confidence clusters.)}
    \label{fig:cluster_detection}
\end{figure}

\begin{figure*}
    \centering
    \begin{minipage}{0.8\linewidth}
    \includegraphics[width=\linewidth]{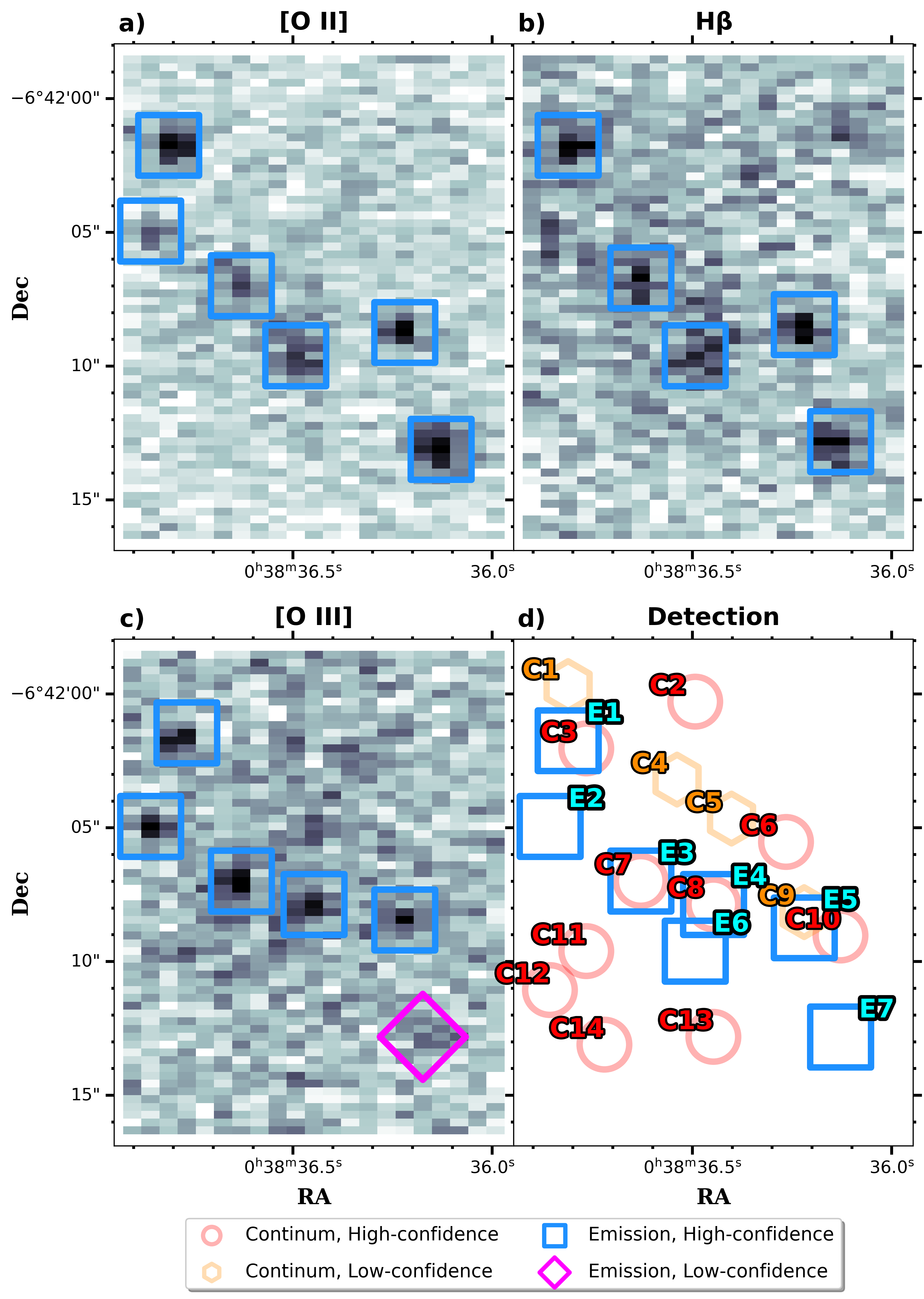}
    \caption{Cluster detections in the emission line slices.
    We present high-confidence 
    and low-confidence 
    detections as defined previously (see text and Figure \ref{fig:cluster_detection}) in each of three emission lines in panels (a)--(c). 
    In panel (d), we plot both the continuum-selected (C) and emission line-selected (E) sources and label them with numbers for identification.
    }
    \label{fig:EC_detection}
    \end{minipage}
\end{figure*}

\begin{deluxetable*}{lllccrrrr}
\tablewidth{0pt}
\tablecaption{Cluster Coordinates and Fluxes \label{table:cluster}}
\tablehead{
 \multicolumn{1}{l}{Name} & \multicolumn{1}{l}{Alt. Name} & \multicolumn{1}{l}{Confidence}& \multicolumn{2}{c} {Coordinates} & 
  \multicolumn{4}{c}{Instrumental Flux [10$^3$ counts]} \\
   & & & RA & Dec & Continuum & [O II] & H$\beta$ & [O III] 
}
\startdata
C1  & ... & L & 0:38:36.8       & $-$6:41:59.7       & 17.3 $\pm$ 2.0 & 0.0 $\pm$ 0.1 & $-$0.1 $\pm$ 0.1 & 0.0 $\pm$ 0.1 \\
C2  & ... & H & 0:38:36.5       & $-$6:42:00.3       & 26.4 $\pm$ 1.9 & 0.0 $\pm$ 0.1 &  0.1 $\pm$ 0.1 & 0.3 $\pm$ 0.1 \\
C3  & E1 & H & 0:38:36.8       & $-$6:42:01.7       & 44.0 $\pm$ 1.9 & 1.2 $\pm$ 0.1 & 1.0 $\pm$ 0.1 & 0.7 $\pm$ 0.1 \\
C4  & ... & L & 0:38:36.5       & $-$6:42:03.2       & 22.0 $\pm$ 1.9 & $-$0.1 $\pm$ 0.1 &  0.1 $\pm$ 0.1 & 0.3 $\pm$ 0.1 \\
C5  & ... & L & 0:38:36.4       & $-$6:42:04.7       & 15.8 $\pm$ 1.9 & 0.2 $\pm$ 0.1 & 0.0 $\pm$ 0.1 & 0.5 $\pm$ 0.1 \\
E2  & ... & H & 0:38:36.9       & $-$6:42:05.0       & -9.3 $\pm$ 1.9 & 0.6 $\pm$ 0.1 & 0.4 $\pm$ 0.1 & 0.9 $\pm$ 0.1 \\
C6  & ... & H & 0:38:36.3       & $-$6:42:05.5       & 30.5 $\pm$ 1.9 & 0.2 $\pm$ 0.1 &  0.2 $\pm$ 0.1 & 0.4 $\pm$ 0.1 \\
C7  & E3 & H & 0:38:36.6       & $-$6:42:07.0       & 58.8 $\pm$ 1.9 & 0.6 $\pm$ 0.1 & 0.7 $\pm$ 0.1 & 1.0 $\pm$ 0.1 \\
C8  & E4, NSC & H & 0:38:36.5   & $-$6:42:07.9      & 215.9 $\pm$ 2.0 & 0.2 $\pm$ 0.1 & 0.2 $\pm$ 0.1 & 0.6 $\pm$ 0.1 \\
C9  & E5 & H & 0:38:36.2       & $-$6:42:08.7       & 39.2 $\pm$ 1.9 & 1.0 $\pm$ 0.1 & 1.2 $\pm$ 0.1 & 0.7 $\pm$ 0.1 \\
C10  & ... & H & 0:38:36.1       & $-$6:42:09.0       & 33.3 $\pm$ 1.9 & 0.4 $\pm$ 0.1 &  0.1 $\pm$ 0.1 & 0.2 $\pm$ 0.1 \\
C11  & ... & H & 0:38:36.8       & $-$6:42:09.6       & 59.7 $\pm$ 1.9 & 0.1 $\pm$ 0.1 &  0.1 $\pm$ 0.1 & 0.3 $\pm$ 0.1 \\
E6  & ... & H & 0:38:36.5       & $-$6:42:09.6       & 16.8 $\pm$ 1.9 & 0.9 $\pm$ 0.1 & 0.8 $\pm$ 0.1 & 0.3 $\pm$ 0.1 \\
C12  & ... & H & 0:38:36.9       & $-$6:42:11.1       & 13.6 $\pm$ 1.9 & 0.1 $\pm$ 0.1 &  0.3 $\pm$ 0.1 & 0.0 $\pm$ 0.1 \\
C13  & ... & H & 0:38:36.4       & $-$6:42:12.8       & 34.5 $\pm$ 1.9 & 0.3 $\pm$ 0.1 &  0.3 $\pm$ 0.1 & 0.0 $\pm$ 0.1 \\
C14 & ... & H & 0:38:36.7       & $-$6:42:13.1       & 23.1 $\pm$ 1.9 & 0.2 $\pm$ 0.1 &  0.1 $\pm$ 0.1 & $-$0.1 $\pm$ 0.1 \\
E7  & ... & H & 0:38:36.1       & $-$6:42:13.1       & 14.1 $\pm$ 1.9 & 1.3 $\pm$ 0.1 & 1.2 $\pm$ 0.1 & 0.6 $\pm$ 0.1 \\
\hline
\enddata
\end{deluxetable*}

\subsubsection{Clusters in the Continuum Image}
\label{sec:finding_cont}

We first detect clusters based on their stellar light alone, without contamination from the principal emission lines. 
We construct the continuum-only slice to avoid potential contamination from [O II], H$\gamma$, H$\beta$, and [O III] by excluding a 5\AA\ wide spectral regions near each of these emission lines. The 5\AA\ corresponds to a spectral regions broader than twice the FWHM of ILB and approximately 300km s$^{-1}$, larger than any recessional velocity offset due to the expected internal rotation or velocity dispersion of the galaxy.
 
Because the clusters are embedded within the diffuse light of the host galaxy, local background variations can affect their detection.
To address this concern as much as possible, we utilize SEP to subtract the spatially varying background and measure the global background noise. We use a small mesh size of 2.04 $\times$ 2.04 arcsec (3$\times$7 spaxels) to subtract the local background. We find that this size is sufficiently large (approximately 2.4 times the FWHM of the PSF) to measure the visible variations in the background.

After background subtraction, we employ SEP again to identify star clusters, which we define as groups of at least three adjacent spaxels, each with a flux $> 2\sigma$ above the background. 
We detect ten objects (marked by red circles in Figure \ref{fig:cluster_detection}), including the NSC. 
We designate these objects as \textit{high-confidence clusters.} 
The requirement that at least three spaxels each have $> 2\sigma$ confidence implies that the overall cluster detection is well above $2\sigma$ confidence, although the spaxel values are not fully independent, so the minimum confidence level is lower than that of three independent $2\sigma$ detections, which would be close to 4$\sigma$.

Next, we mask the previously identified clusters and repeat the process to detect additional clusters that may have been obscured. We use a lower threshold, where the flux in each spaxel needs only to be $> 1.5\sigma$ above the background, and identify five additional clusters. 
SEP provides two estimates of the central position.
We reject the one detection where the separation between the two estimates of the center exceeds the PSF FWHM, which we interpret to show that the detection is highly contaminated by the remnant of the nearby, previously identified high-confidence cluster. 
We designate the newly identified four objects as \textit{low-confidence clusters,} and they 
are highlighted with orange hexagons in Figure \ref{fig:cluster_detection}. 
Despite the lower formal confidence, these still appear visually convincing. Furthermore, since each cluster consists of at least three spaxels, their total significance corresponds to approximately a $2.6\sigma$ detection or more.
The union of high and low confidence clusters comprises our cluster sample.

We explored the dependence of the detection procedure to our choice of background mesh size. The size we adopted produced results most consistent with our visual selection and an effort to minimize false detections and maximize completeness. For comparison, doubling the mesh size results in 6 and 4 high- and low-confidence clusters, respectively, rather than 10 and 4. The reader can judge our results using Figure \ref{fig:cluster_detection}. There are suggestions of additional clusters that we have not identified, but the identified clusters appear to be robust detections. However, we do not expect this work to be the final word on the cluster population in inner regions of the Disco Ball.

We present the coordinates and instrumental aperture fluxes of the 14 clusters identified from the continuum image in Table \ref{table:cluster}. We calculate the flux of each cluster by first summing the spaxel values within a radius equal to the PSF's FWHM and subtracting the local background. 
To estimate the background, we measure the flux within an annular region with inner and outer radii of 1.5 and 3.5 FWHM centered on each object, excluding areas associated with any neighboring clusters.
We will estimate the corresponding magnitudes of these objects in \S\ref{sec:results}.

\subsubsection{Clusters in the Emission Slices}
\label{sec:finding_emission}
Some clusters are strongly line-emitting, which is indicative of ongoing or very recent ($\lesssim$ 20Myr) star formation. Although these are not what would typically be referred to as GCs, detecting sources in the emission line slices provides information about star formation rates, the ionization state of the gas, and the local gas kinematics.

We reconstruct emission line slices based on the bandpasses of interest ([O II]3727, H$\beta$, and [O III]5007). For each slice, we use a bandwidth of 5\AA, corresponding to about 2 times the FWHM of the ILB, centered at the redshifted wavelength of each of the three emission lines. To isolate the emission line flux from the local continuum, we subtract the continuum, measured using slices that are centered at wavelengths that are 5\AA\ shorter and longer.

We apply the same detection methodology described in \S\ref{sec:finding_cont}, and show the results in Figure \ref{fig:EC_detection}. Panels (a)--(c) display the detection results in each emission line slice, and Panel (d) presents all of the identified emission line clusters, numbered from highest to lowest in declination (E+number), regardless of their statistical significance. We also mark and number all of the identified clusters in the continuum image (C+number). Note that some clusters have both a C and an E label. We present the measured properties of the identified sources in Table \ref{table:cluster}.

In the [O II] slice (Panel a), we identify six high-confidence clusters, marked by blue squares. Among these, E1, E3, and E5 have clear counterparts in the continuum image, whereas E2, E6, and E7 lack obvious continuum counterparts.
The NSC (C8 or E4) is not detected in [O II] emission, but there is an [O II] emitting clump (E6) near the NSC that has no counterpart in the continuum image. This source has a corresponding detection in the H$\beta$ slice (Panel b) but shows only weak [O III] emission, falling below our detection threshold.

In the [O III] slice (Panel c) we detect E1, E2, E3, E5, and E7, which were all also detected in [O II]. However, one notable difference is that the NSC has a corresponding strong [O III] emission feature (E4) that is weak or absent in the [O II] and H$\beta$ slices. This aspect will be discussed further in \S\ref{sec:o2o3}. Despite variations in line strength across the slices, all emission line sources are confidently detected in at least one slice.

We present the coordinates and instrumental aperture line fluxes of the emission line clusters in Table \ref{table:cluster}. Again, the instrumental fluxes are calculated by summing all of the pixel values within the given, continuum-subtracted bandpass and the spaxels within a radius of one PSF FWHM.

\subsection{Spectral analysis}
\label{sec:spec}

\begin{figure*}
	\includegraphics[width=\linewidth]{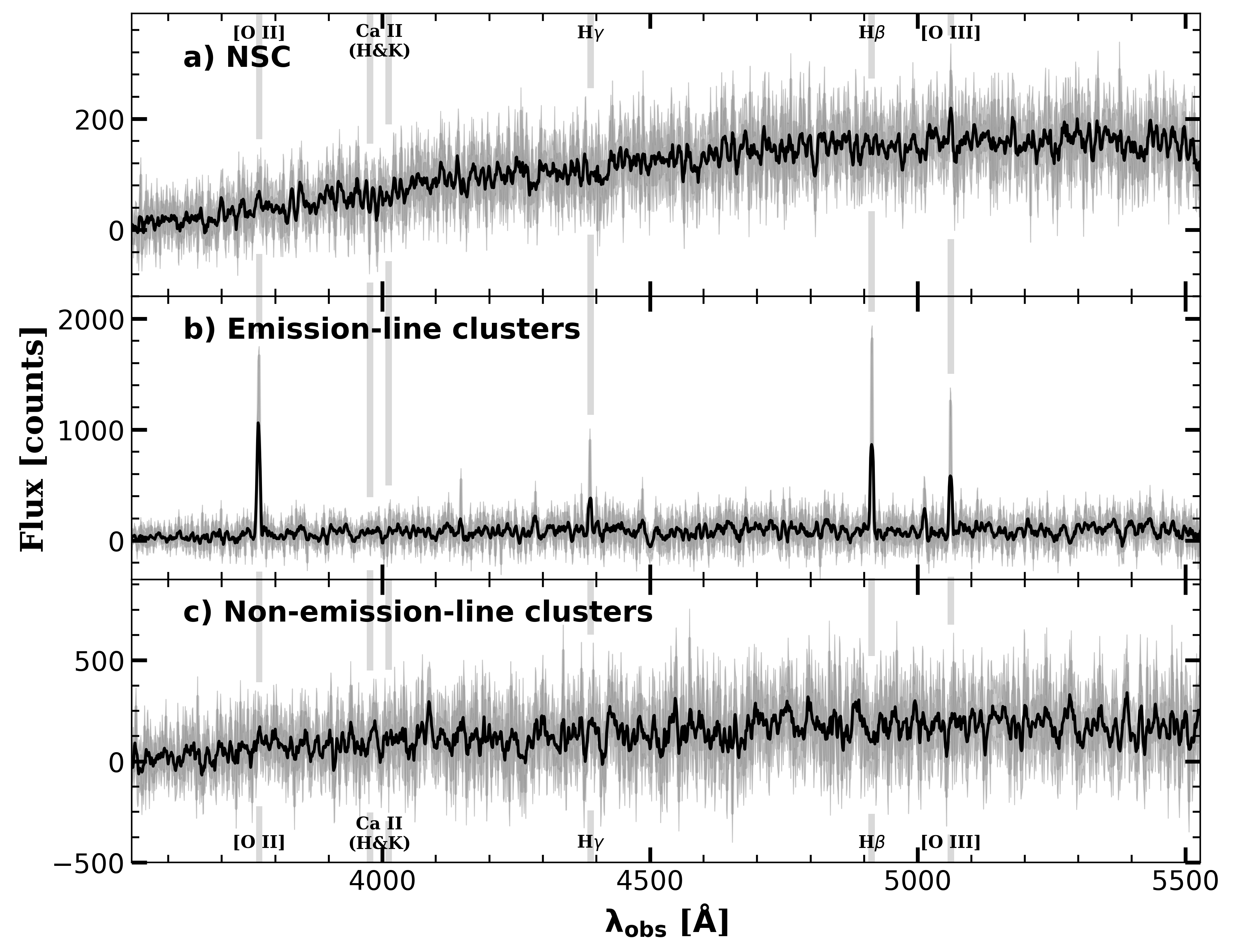}
    \caption{The extracted spectra for (a) the NSC, (b) emission-line clusters, and (c) clusters without emission lines. We identify common spectral features identified for reference and lightly shaded regions indicate the spectral uncertainties.
    }
    \label{fig:spec}
\end{figure*}

In this subsection, we extract different sets of spectra from the KCWI data to explore the kinematics of different galactic components. 
To characterize the diffuse light of the galaxy, we stack all the spaxels within 0.5 $r_{\rm e}$ of the galaxy and mask all of the identified clusters using an aperture with a radius of one PSF FWHM.
To extract individual cluster spectra, we select all spaxels within a circular aperture with a radius of one PSF FWHM radius centered on each cluster. To estimate and subtract the host galaxy's contribution to the spectra of any cluster, we use the same annular background region employed in the flux measurements (\S \ref{sec:finding_cont}). 
We show the resulting NSC spectrum in Figure \ref{fig:spec}-(a), 
the stacked spectrum of emission clusters other than the NSC in Panel (b), and the stacked spectrum of clusters identified in the continuum image that are undetected in the emission line slices in Panel (c). Due to S/N concerns, we only consider the NSC and the emission line clusters individually in subsequent analyses. For the non-emission-line clusters, we focus solely on the stacked spectrum.

\subsubsection{Velocity and velocity dispersion measurements}
\label{sec:ppxf}

We utilize the Penalized PiXel-Fitting algorithms \citep[pPXF, see][for details]{ppxf} to derive the bulk stellar and gas kinematics. The fitting includes three independent spectral components: the stellar continuum, Balmer emission lines, and oxygen emission lines.
For modeling the stellar population, we use the X-shooter Spectral Library \citep[XSL, see][]{xsl}. We allow pPXF to apply a fourth-order additive Legendre polynomial to correct the template continuum shape during the fit. We adopt the measured ILB and set the initial values for the velocity and velocity dispersion at 3250 and 30km s$^{-1}$, respectively.

We apply pPXF to the extracted spectra of the host galaxy (within 0.5$r_{\rm e}$ of the galaxy, masking all identified clusters), the NSC, the individual emission line clusters, and the stacked continuum-only clusters (See, \S\ref{sec:spec}).
We reject a result if the resulting peak flux of the best-fit model is lower than the noise in the continuum. We also consider meaningless, and reject, any individual measured value if it is smaller than its associated uncertainty. We present our results in Table \ref{table:velocity_sigma}. Rejected values are represented with ellipsis.

For the host galaxy spectrum, we are able to measure the recessional velocity independently with each of the three spectral components and the results are consistent within the uncertainties. The resulting values for the velocity dispersion are also consistent among the two spectral components for which we recover a measurement, but caution is warranted given that the values are smaller than the ILB. Although it is mathematically possible to measure a kinematic broadening that is smaller than the instrumental broadening, such results are often highly dependent on S/N and critically sensitive to the adopted ILB. At this point in our presentation, we simply retain these values for further consideration and note the agreement between the two values.

For the NSC spectrum, we are able to measure the recessional velocity using either the stellar spectral component or the oxygen lines.  
The resulting values are consistent within the uncertainties and confirm that the central point source is associated with the host galaxy. 
We also obtain a measure of the velocity dispersion of the NSC.
Although in this case the kinematic broadening is greater than the ILB, it comes with a large uncertainty and is therefore also suspect. This value is also more difficult to interpret because gas kinematics is susceptible to a wider range of influences. We will return to this topic in \S\ref{sec:nsc}. 

For the clusters, we obtain mixed results. We are unable to measure a velocity from the stacked continuum-only spectra. 
The situation is more positive for the emission line clusters where we are able to measure individual recessional velocities for the seven emission line clusters.

\begin{deluxetable}{llrr} 
\tablecaption{Kinematics of the host galaxy and NSC
\label{table:velocity_sigma}}
\tablehead{
Object & Component & \multicolumn{1}{c}{$cz$}  & \multicolumn{1}{c}{$\sigma_v$} \\
&& \multicolumn{1}{c}{[km s$^{1}$]} & \multicolumn{1}{c}{[km s$^{1}$]} 
}
\startdata
Host galaxy & Stellar & 3266 $\pm$ \ 6 & 50 $\pm$ 10 \\
Host galaxy & Balmer & 3262 $\pm$  \ 7& ...  \\
Host galaxy & Oxygen & 3273 $\pm$ \ 7 & ...  \\
NSC (E4)  & Stellar & 3257 $\pm$ 22 & 51 $\pm$ 44 \\
NSC (E4) & Oxygen & 3280 $\pm$ 51 & 113 $\pm$ 54 \\ 
Continuum Clusters & ... & ...  & ... \\  
\enddata
\end{deluxetable}

\begin{deluxetable}{ccc}
\tablecaption{Emission line cluster recessional velocities \label{table:velocity_sigma_ec}}
\tablehead{
Cluster & \multicolumn{1}{c}{Balmer Line $cz$} & \multicolumn{1}{c}{Oxygen Line $cz$} \\
& \multicolumn{1}{c}{[km s$^{-1}$]} & \multicolumn{1}{c}{[km s$^{-1}$]} 
}
\startdata
E1 & 3244 $\pm$ \ 7 & 3252 $\pm$ \ 7 \\
E2 & 3265 $\pm$ 23  & 3252 $\pm$ \ 8  \\
E3 & 3248 $\pm$ \ 8 & 3268 $\pm$ 11 \\
E5 & 3264 $\pm$ \ 6 & 3276 $\pm$ \ 8 \\
E6 & 3261 $\pm$ 10  & 3254 $\pm$ 12\\
E7 & 3267 $\pm$ \ 7 & 3264 $\pm$ \ 6 \\
\enddata
\end{deluxetable}

\section{Results}
\label{sec:results}

\subsection{The Cluster Population}

We now present an estimate of the size of the overall cluster population, and use this result to estimate the total galaxy mass through the scaling relation between the number of clusters and mass, which has been shown to extend to this mass regime \citep{forbes+18,Toloba_2018,zaritsky_2022}. We discuss whether the identified clusters, both continuum-only and emission line clusters, are associated with the host galaxy and assess whether they are similar to cluster populations observed in other galaxies. After that, accounting for completeness, we estimate the total number of possible GCs in the system and derive a corresponding galaxy mass.

\subsubsection{Membership}

Given the agreement in recessional velocities (all within 20km s$^{-1}$), we confirm that all seven emission-line clusters, including the NSC, are associated with the host galaxy. 
However, because of the low S/N ratio, we are unable to measure recessional velocity even for the stacked continuum-only clusters, leaving their association with the host galaxy unconfirmed through kinematic analysis.
Nevertheless, there are other reasons to support the association: (1) the localized overabundance of such sources in the KCWI image projected onto the host is unmatched in our KCWI observations of 44 candidate UDGs, (2) the continuum-selected clusters that also happen to have emission lines are confirmed to be members, and (3) none of the sources are confirmed to be background sources. To expand on the first point, many of the KCWI images show no point sources superposed on the host and none show as large a number as seen in the Disco Ball. To expand on the last point, we examined the spectrum of each individual cluster and found none with spectral features corresponding to a different redshift.

\subsubsection{The Nature of the Detected Sources}

Even if all of the detected sources are associated with the host galaxy, they may not be stellar clusters, and then they may not be GCs. To examine the nature of the detected sources we take two steps.
First, we exclude the NSC from this analysis because NSCs are typically significantly more massive than individual GCs \citep{neumayer} and so not part of a potential GC population. 
We will discuss the NSC separately in \S \ref{sec:nsc} and what the presence of an NSC may mean for the GC population in \S \ref{sec:discussion_Ngc}.
Second, we exclude sources with emission lines. In one case there is no continuum source detected and, even when there is, these are evidently young stellar populations that will fade significantly as they evolve. To provide intuition on the possible degree of fading, we consider a metal-poor, [Fe/H]$ = -1$, simple stellar population of 100Myr, already well older than what would produce emission lines. 
Using Prospector \citep{prospector} with the FSPS stellar populations model \citep{fsps1, fsps2}, we find that this population will fade by 3.5 mag in 10Gyr. Therefore, the degree of fading until the emission line clusters become ancient clusters must be $>$ 3.5 mag. Even if they survive their youth, which is not necessarily the case for many young stellar clusters \citep[e.g.,][]{chandar}, they would fade well below the globular cluster luminosity function (GCLF) peak and so not be part of a population that is typically observed in UDGs \citep[cf.][]{Amorisco2016,Beasley_2016,lim,Saifollahi+2022,marleau}. We exclude these objects from further consideration as GCs.

\begin{figure}
    \includegraphics[width=\columnwidth]{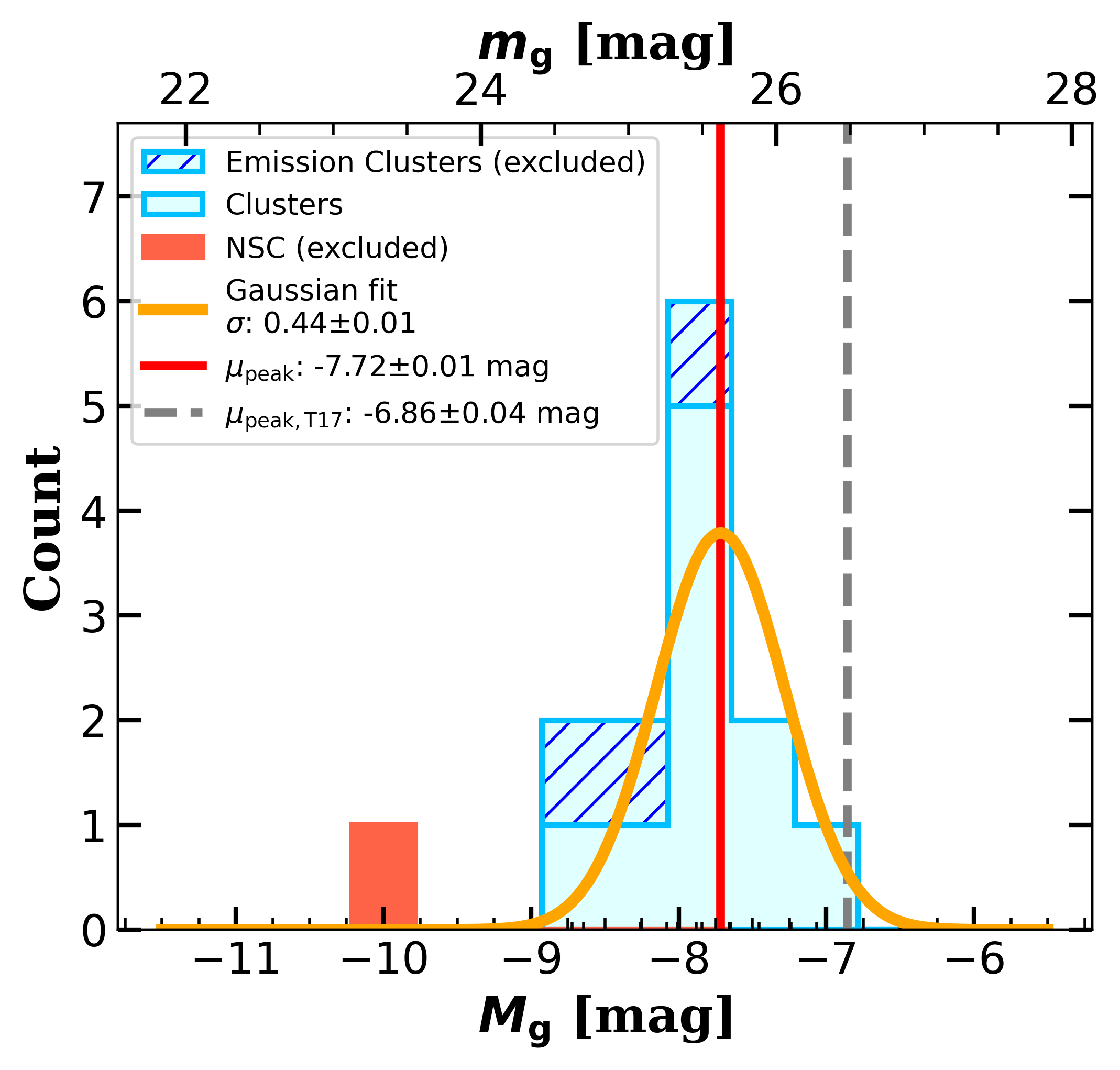}
    \caption{The luminosity function (LF) of detected sources. The NSC, indicated in red, is excluded from further analysis of the population. Additionally, clusters with line emissions, highlighted in the hashed histogram, are also excluded. A Gaussian fit to the remaining population is shown in orange and the peak of the observed LF is shown in red. For comparison, we also present the location of the GCLF peak from \cite{Taylor+2017} as the gray dashed line. }
    \label{fig:GCLF}
\end{figure}

We now assess the nature of the remaining sources by estimating their total numbers and measuring their luminosity function (LF). If both of these measures are consistent with the properties of GC systems in other UDGs \citep[e.g.][]{beasley2016,Toloba_2018,Saifollahi+2022,marleau}, then we gain confidence in identifying these as GCs. We start by considering the LF, because that leads to issues of completeness that are necessary to address prior to determining the size of the population.

We measure the cluster LF and compare it to the canonical GCLF, which is known to have a near-universal peak or turnover \citep{Jacoby+1992, McLaughlin+2005, Brodie+2006}, even for clusters in UDGs \citep{Saifollahi+2022,marleau}.
We use the PYPHOT package \citep{zenodopyphot} to measure the flux within the SDSS $g$-band filter bandpass using the spectra. We convert the flux to a $g$-band magnitude using the previously measured $g$-band magnitude of the NSC \citep{khim+24_nsc}.

We present the resulting LF in Figure \ref{fig:GCLF}, where the peak magnitude of the Gaussian fit is $-7.7$ mag, approximately 0.8 magnitude brighter than the typical peak magnitude of $\sim-6.9$ \citep[cf.][]{Taylor+2017}.
Our sample is clearly not a direct analog to a standard GC population. Even excluding the emission line clusters, which are certain to fade significantly, it is clear that our clusters are more luminous than the typical GC. We present four possible interpretations. 

\medskip
\noindent
$\bullet$ Photometric incompleteness. 
If our sample suffers from significant incompleteness, the true peak of the cluster population could be substantially fainter than our measurement. This scenario would imply a much larger underlying cluster population.
To examine this possibility, we first construct cumulative nonparametric Epanechnikov-kernel probability density estimates of our LF and that of \cite{Taylor+2017}. Under the assumption that the LF for the Disco Ball GCs has the same shape as the canonical GCLF, then up to the point where we are complete we should find the same fraction of the total number of GCs. 
We then correct for the missing clusters fainter than this, using the GCLF from \cite{Taylor+2017}, assuming that we are complete up to our peak absolute magnitude  ($M_g = -7.7$ mag). We infer that in this scenario we are missing about 60\% of the clusters. We will return later to discuss whether this leads to an estimated total number of GCs that invalidates this possibility.

\medskip
\noindent
$\bullet$
Non-standard GCLF. Various studies \citep{Shen_2021,janssens,romanowsky,tang} have identified candidate GC populations in low surface brightness galaxies that are overly luminous relative to the canonical GCLF. This trend may even be a rather general phenomenon among low surface brightness galaxies \citep{li}. If so, then the most conservative assumption, in terms of minimizing $N_{\rm GC}$, is that we are finding a similar population and that we are complete in our cluster identification.

\medskip
\noindent
$\bullet$ Fading.
The presence of emission-line clusters indicates that a significant fraction of our cluster population is young and will fade significantly over time. 
As we described previously, a 100Myr simple stellar population will fade 3.5 mag over 10Gyr, far more than needed to align our measured LF peak with the canonical one. Once we have accepted that cluster formation did not cease entirely $\sim10$Gyr ago, we must accept the possibility that some clusters may be of intermediate age. For consideration,
$\sim$4Gyr clusters will fade by $\sim$0.75 mag in the $r$-band by the time they are 10Gyr old.
As such, we can posit that the observed population consists of intermediate-age clusters that will fade and then match the canonical GCLF. 
However, it remains unclear if these clusters formed continuously over time or in specific episodes. Another UDG previously identified as hosting overluminous clusters, DGSAT I, also shows evidence of recent star formation activity despite its otherwise quiescent appearance \citep{janssens}.
The existence of intermediate age clusters, in general, is not novel for dwarf galaxies; see, for example, the Large Magellanic Cloud \citep{sarajedini}, but it is not generally considered a possibility in UDGs.

\medskip
\noindent
$\bullet$ Disruption.
If these sources are not GCs and one does not invoke fading, then one must eliminate these objects in another manner because compact objects of this luminosity do not exist in galaxies unless they are GCs. Clusters are susceptible to a variety of disruptive phenomena across time \citep[cf.][]{gieles_2011,webb_2024}. 
Of course this scenario, like the fading scenario, requires that the clusters are young enough to allow for subsequent disruption, and therefore fading would also be integral to making these objects disappear from view.
In such a scenario, the observed LF could be a combination of younger clusters, that will be disrupted and/or fade to alleviate the disagreement with the canonical GCLF, and a set of older clusters that satisfy the GCLF. 

\medskip
The four scenarios provide a wide range of possibilities spanning from there being many times as many GCs as clusters observed to there being no ancient GCs or clusters that will evolve to what we would consider ancient GCs. Until age constraints are obtained for the continuum-only clusters, this level of uncertainty will remain. Deeper imaging will  help us determine if there is already a GC population at and below the canonical GCLF turnover magnitude. However, as we discuss further below, this level of uncertainty plagues most measurements of GC populations in UDGs.

\subsubsection{Radial Completeness}

The limited field of view, which extends only to $\sim$ 0.6 effective radii, also limits us from finding all of the clusters. To correct for radial incompleteness, we assume that the spatial distribution of clusters follows that of the stars, consistent with findings in other UDGs \citep{Saifollahi+2022,marleau}. 
We model the stellar light profile using a S\'ersic profile, adopting parameters from \cite{khim+24_nsc}: a S\'ersic index of 0.91, an effective radius of 15.0 arcsecs, and an axis ratio of 0.75. Of the 14, nine of our clusters fall within the 0.6$r_{\rm e}$ isophote. We correct the cluster counts within 0.6$r_{\rm e}$ by a factor of 3.8, which is derived from the ratio of unresolved light within  0.6$r_{\rm e}$ to the total luminosity, to obtain an estimate of the radially complete population.  We note that other studies \citep[see][for discussion]{forbes24} adopt larger radial corrections because they posit that the GCs are less concentrated than the host galaxy stars. Our approach is again conservative in limiting $N_{\rm GC}$.

\subsubsection{Number of clusters and halo mass}
\label{sec:Ngc_halomass}

Following the discussion of the cluster sample, the LF, and radial completeness, we now present estimates of the total number of clusters in the Disco Ball, acknowledging the high level of uncertainty. By excluding the emission-line clusters and assuming that the non-emission line clusters fade moderately and are not disrupted, these objects would be identified as GCs in a typical study. We associate their number with $N_{\rm GC}$, consistent with definitions in previous work. Although this is not the sole possible interpretation of these sources, similar uncertainties exist in most studies of GCs in UDGs (\S\ref{sec:discussion_Ngc}).

In the interest of specificity, on the low end of the range of estimates, it is possible to dismiss the entire sample of objects by postulating that they will fade beyond recognition or be disrupted. As such, the lower limit on $N_{\rm GC}$ is zero. However, we dismiss this as a possibility because UDGs of similar size and luminosity tend to have several tens of clusters \citep{2017vanDokkum,Toloba_2018, Gannon2022, Saifollahi+2022}. 

On the high end of the range, we can assume that the discrepancy between the LFs is solely due to photometric incompleteness.
If so, we estimate the system contains 115 clusters after accounting for both the photometric (factor of $\sim$3.3) and radial (factor of $\sim$3.8) incompleteness described earlier. This estimate is significantly larger than for similar UDGs \citep{Gannon2022,Saifollahi+2022,Toloba_2023}. As such, we conclude that our measurement of a brighter GCLF turnover is likely not due to gross incompleteness.

As we have noted previously, our preferred value for $N_{\rm GC}$ comes from adopting the scenario where the current LF fades to match the canonical GCLF and we only need to apply the correction for radial incompleteness. In this case, $N_{\rm GC}$ = 34 $\pm$ 11 (adopting Poissonian uncertainty), more in line with expectations drawn from previous studies of similar UDGs. We would reach a similar conclusion if instead we posited that the GCs in the Disco Ball were simply overluminous.

To estimate the Disco Ball's halo mass, we use the scaling relation between $N_{\rm GC}$ and mass, as applied in \cite{Saifollahi+2022}. Those authors used the scaling relation from \cite{Harris2017}, adopted an average GC mass of 2 $\times$ 10$^{5}$\Msol, and derived 
\begin{linenomath}
    \begin{equation}
        M_{\mathrm{h}} = 5.12 \times 10^{9} \times N_{\mathrm{GC}} \, \mathrm{M_\odot}. 
        \label{eq:scaling_relation}
    \end{equation}
\end{linenomath}
For our preferred value of $N_{\rm GC}$, we estimate the halo mass of Disco Ball to be $M_{\rm h} = 10^{11.25 \pm 0.26}$\Msol. 
For context, \cite{Saifollahi+2022}, using their measurement of $N_{\rm GC}$ for DF44 and the same scaling relation we applied, estimated $M_{\rm h} = 10^{11.01}$\Msol, comparable but somewhat smaller than the Disco Ball halo mass. On the other hand, \cite{gannon24} find $N_{\rm GC}=74$ for DF44, and correspondingly that $M_{\rm h} = 10^{11.58}$\Msol, which is somewhat larger than what we find for the Disco Ball.
We will compare our $N_{\rm GC}$-based estimate of $M_{\rm h}$ for the Disco Ball to dynamical estimates in \S\ref{sec:rot_halomass}.

\subsection{Galaxy Kinematics}

\subsubsection{Rotation}

The indication of distributed star formation that we have found in the Disco Ball suggests the presence of a cold gaseous disk, which would then likely imply rotation. 
H{\small I}-rich UDGs \citep{leisman} often show rotation \citep{mancera,karunakaran}, and although we do not know the H{\small I} properties of the Disco Ball, the ongoing or recent star formation testifies to the recent presence of gas.

\begin{figure}
    \includegraphics[width=\columnwidth]{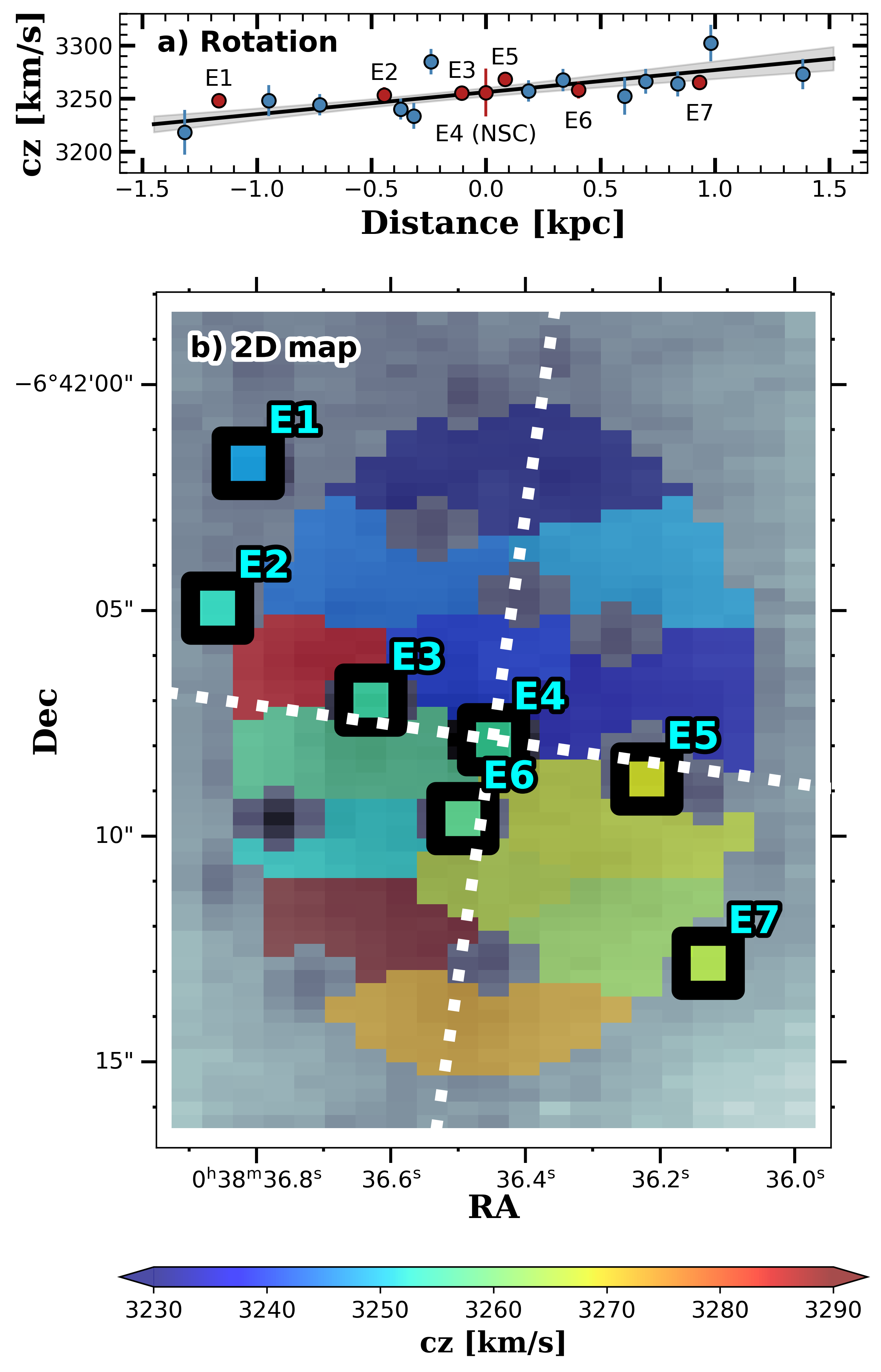}
    \caption{\textit{Panel (a)}: The recessional velocity measured in bins (blue) and for emission line clusters (red) as a function of position along the photometric major axis. For the clusters, we use the mean recessional velocity derived from their emission lines, whereas we use the continuum velocity for the NSC (E4) and for the diffuse light within the bins. The black solid line and shaded area represent the best fit linear rotation curve and its 1$\sigma$ error, respectively.  
    \textit{Panel (b)}: The recessional velocity of the galaxy and clusters shown in 2D. The white dotted lines represent the photometric minor, or rotation, axis (PA =  82$^\circ$) and photometric major axis (PA $= -8^\circ$). The median uncertainty in the measured velocities is $\sim$ 12km sec$^{-1}$.}
    \label{fig:rotation}
\end{figure}

To determine the bulk internal kinematic properties of the Disco Ball we measure the spatially-resolved recessional velocity of the diffuse galaxy light. We first mask all of the identified clusters (Figure \ref{fig:EC_detection}). 
We then divide the data within 0.5$r_{\rm e}$ into bins using Voronoi 2D binning, targeting a velocity uncertainty of less than 20km sec$^{-1}$ per bin. This approach results in a minimum S/N of 200 per bin and a total of 13 bins.
For each bin, we extract a spectrum of the diffuse light and measure the recessional velocity (Figure \ref{fig:rotation}). The velocities for each bin are derived from their absorption line spectra and have an uncertainty of approximately 12km sec$^{-1}$. 

We plot the recessional velocities, against distance from the galaxy center along the photometric major axis (PA = $-8^\circ$) in Figure \ref{fig:rotation}-(a). We perform a simple Spearman correlation test, showing that the trend is statistically significant with a confidence level $>$ 3$\sigma$ (p-value of 0.0021). The best-fit line (black solid line) and its uncertainty (shaded region) suggest that the galaxy has a projected rotation velocity of 35.0 $\pm$ 6.7km s$^{-1}$ at 0.5$r_{\rm e}$. Given the nature of the data, we cannot distinguish between a purely linear rotation curve over these radii and one that reaches a turnover and flattens. 

The emission line clusters, which are aligned roughly from northeast to southwest in the galaxy (Figure \ref{fig:EC_detection}), have individual velocities that are consistent with those measured for the main galaxy body. This agreement shows both that the clusters are indeed embedded in the body of galaxy and that our velocity measurements for both the individual clusters and the diffuse light within bins are generally precise to $\lesssim 10$km sec$^{-1}$.

If we assume a thin disk geometry, the axis ratio (b/a = 0.75) implies an inclination of 41$^\circ$ and that the deprojected rotation velocity at 0.5$r_{\rm e}$ is 53.0 $\pm$ 10.2km sec$^{-1}$. The agreement between the rotation gradient and the photometric major axis fits this scenario well, but the system could still be a rotating oblate system, like a low-mass galaxy bulge, in which our inclination correction based on the axis ratio would be incorrect. Even there, we would expect there to be some inclination correction to the rotation velocity and significant velocity dispersion, both of which would have to be accounted for in any calculation of this system's dynamical support.

\subsubsection{Estimating the Halo Mass}
\label{sec:rot_halomass}
UDGs are generally assumed to be pressure-supported systems because of their spheroidal morphology. Among measurements of the stellar populations \citep[e.g.][]{vdk19} 
there is often no evidence of significant rotation. 
In contrast, H{\small I}-rich systems do show rotation \citep{mancera,karunakaran}, although there are disagreements on how well the systems satisfy the Tully-Fisher relation. The Disco Ball is the rare UDG with rotation derived from the diffuse stellar population.

The rotation velocity, $v_{\rm rot}$, of this galaxy provides a relatively simple way to estimate the mass within radius $r$,
\begin{linenomath}
    \begin{equation}
        M = r v_{\rm rot}^2 / (A G ),
        \label{eq:rotation_formula}
    \end{equation}
\end{linenomath}
where $A$ is a correction factor that accounts for deviations from a purely spherical mass distribution and $G$ is the gravitational constant. We refer to this equation as the rotation formula.
For a series of eight local disk galaxies, \cite{feng} find an average value of $A = 1.66$, but those are galaxies that are baryon-dominated within $r_{\rm e}$. Instead, we will neglect the influence of the stellar disk and adopt $A=1$ for the Disco Ball, but note that by doing so we are overestimating the mass to whatever degree that A is greater than 1.
We find an enclosed mass at $r_{\rm e}$ of $M_{\rm e} =10^{9.3 \pm 0.2}$\Msol, assuming a flat rotation curve between 0.5$r_{\rm e}$ and $r_{\rm e}$. 
For comparison, the scaling relation used by \cite{2023MNRAS.519..871Z} to estimate the velocity dispersions and masses of galaxies suggests that this system has an equivalent $v_{\rm rot}$ of 47 $\pm$ 11km sec$^{-1}$ and $M_{\rm e} = 10^{9.5 \pm 0.2}$\Msol. Additionally, the Wolf mass estimator \citep{wolf+10} suggests an enclosed mass at $r_{\rm e}$ of $M_{\rm e} = 10^{9.6 \pm 0.2}$\Msol.
The close agreement is reassuring given the many assumptions and uncertainties in mass estimation. 

We estimate the galaxy's halo mass by extrapolating from $M_{\rm e}$ to $M_{\rm h}$, following the method described in \cite{2023MNRAS.519..871Z}. 
The enclosed mass derived from the rotation formula (Equation \ref{eq:rotation_formula}) suggests a halo mass of $M_{\rm h}=10^{10.3\pm0.6}$\Msol, where we set the uncertainty by propagating the uncertainty in our calculated value of $M_{\rm e}$ because this results in uncertainty that exceeds the scatter in the stellar mass-halo mass relation.
The halo mass extrapolated from the Wolf mass suggests $M_{\rm h} = 10^{10.9 \pm 0.6}$\Msol.
Both of these mass estimates are $< 1.5\sigma$ different than the halo mass derived from $N_{\rm GC}$, which is $M_{\rm h} = 10^{11.25 \pm 0.26}$\Msol, and they are consistent with each other. 
Considering that we have assumed no increase in rotation velocity beyond 0.5$r_{\rm e}$ and that the uncertainty in the mass obtained using $N_{\rm GC}$ only represents the internal statistical uncertainty, we consider these three estimates of $M_{\rm h}$ to be consistent and the scatter among them to be indicative of the uncertainties in determining $M_{\rm h}$. Taking into consideration all three estimates, we conclude that $M_{\rm h} = 10^{11.1\pm0.2}$\Msol. Extending the rotation curve beyond 0.5$r_{\rm e}$ is a key next step in solidifying the dynamical mass estimates.

\subsubsection{Stellar Mass and Baryonic Tully-Fisher Relation}

The Baryonic Tully-Fisher Relation (BTFR) shows the empirical correlation between the rotational velocity of galaxies and their baryonic mass \citep[e.g.,][]{McGaugh+00, 2001ApJ...550..212B, 2001ApJ...563..694V, 2006ApJ...653..240G}. 
The applicability of the BTFR to UDGs remains uncertain, especially for \hi-rich UDGs \citep[e.g.,][]{2020MNRAS.495.3636M, 2023ApJ...947L...9H}. 

The Disco Ball, with its stellar rotational motion, represents a rare case among UDGs because most appear to be pressure-supported systems unless they are strongly star-forming. \citet{Stark+09} and \cite{Trachternach+09} extended the BTFR of \citet{McGaugh+05} with independent samples of low-mass, gas-rich galaxies, showing that the relation holds across a wide range of masses. 
Using the scaling relation from \cite{Trachternach+09}, which closely resembles that of \citet{Stark+09}, we derive a baryonic mass for the Disco Ball of $10^{8.6 \pm 0.3}$\Msol\ for the Disco Ball, assuming that the measured rotation velocity at 0.5$r_{\rm e}$ represents the maximum rotational velocity. This compares well with a measurement of the galaxy's stellar mass of $10^{8.5 \pm 0.2}$\Msol\ , estimated using color-dependent mass-to-light ratios from \cite{roediger} and the measured luminosity and color, as described in \cite{khim+24_nsc}.

The comparison suggests that the Disco Ball is unlikely to be a gas-rich UDG, though its evidence of recent or ongoing star formation indicates the presence of some gas.
In this interpretation, the Disco Ball satisfies the BTFR. An alternative interpretation is that the Disco Ball has a large, yet unmeasured, gaseous baryonic component and that it lies off the relationship.  H{\small I} observations of the Disco Ball will resolve this ambiguity.

\subsection{The NSC}
\label{sec:nsc}

NSCs are interesting for several reasons, but one key aspect is the speculation that they may host intermediate mass black holes (IMBH) \citep{pz,antonini}. This is an active area of investigation. For example, \cite{aravidan} finds evidence for active black holes in NSCs in dwarf galaxies using mid-infrared variability. We investigate whether there is any indication of an active central IMBH in the Disco Ball's NSC.

\subsubsection{Kinematics}

We first look for unusual central gas kinematics.
We measure the velocity dispersion of the gas in and surrounding the NSC by analyzing the line broadening of the [O III] emission line, our strongest line. We fit a Gaussian to the emission line, resulting in a measure of $\sigma$ of 2.17 $\pm$ 0.70\AA, which corresponds to a velocity dispersion of 109.56 $\pm$ 39.09km s$^{-1}$. After subtracting the instrumental velocity dispersion in quadrature, the adjusted velocity dispersion is 130 $\pm$ 48km s$^{-1}$. 
The measurement is highly uncertain, being consistent with zero velocity dispersion at the $\sim 2.5\sigma$ level, but does hint at a larger dispersion than that of the host galaxy or any individual stellar emission line cluster. 
This measurement is consistent with the pPFX result, which measured $\sigma_v = 113 \pm 54$km s$^{-1}$ from the oxygen lines. 
We conclude that the gas kinematics may hint at interesting physics near the nucleus, but that a higher precision measurement is needed before reaching any firm conclusion.

\subsubsection{Emission Line Ratios}
\label{sec:o2o3}

As already mentioned (\S\ref{sec:finding_emission}), a distinguishing characteristic of the NSC is its unusual [O III] emission line strength relative to [O II]. The measured line ratio for the NSC is $3.1 \pm 1.9$, exceeding the average of the ratio, $0.9\pm 0.1$, of the emission line clusters excluding the NSC. Again, a final conclusion requires a higher precision measurement because the difference in these measurements is only significant at $\sim 1.2\sigma.$ Nevertheless, both the kinematics and the emission line ratio support further investigation of the NSC as a potential host of a weakly active nuclear source.

\section{Discussion}
\label{sec:discussion}

\subsection{How Unusual Is the Disco Ball?}

Lessons learned from a detailed study of this one UDG depend on whether the Disco Ball is an example of a class of UDG or a unique outlier. The
Disco Ball is a relatively blue and bright UDG with an NSC, but well within the range of properties of other red-sequence UDGs (Figure \ref{fig:cmd}). The emission lines in the integrated spectrum are weak and would not be identified in a low S/N spectrum. It contains a few narrow emission-line clusters, indicating recent star formation activity, which might be unusual, but this is a feature that is difficult to identify in the original survey imaging used to construct the SMUDGes catalog. Nothing in the original survey data or in what would be a survey-level spectroscopic survey would identify the Disco Ball as an unusual UDG. \cite{kadowaki17} did find UDGs with weak emission lines. As such, we conclude that the full SMUDGes catalog, and other UDG samples, may contain analogs of the Disco Ball.

Following our full investigation, where we are able to measure host galaxy physical parameters such as size, luminosity, $N_{\rm GC}$, $M_{\rm e}$ and estimate $M_{\rm h}$, we again find that the Disco Ball is not unique in any aspect. We may have caught the Disco Ball at a special time in its evolution, where it is actively forming stellar clumps, but otherwise, the galaxy appears to be broadly similar to other UDGs in terms of its size, structure, stellar mass, $N_{\rm GC}$, and $M_{\rm h}$.
We will investigate the KCWI sample for additional examples, although as stated in \S\ref{sec:intro}, we selected this galaxy for further study because it stood out as unusual in our visual examination of the KCWI data cubes. Even so, there are others in that data set that show ongoing or recent star formation \citep{loraine}.

\subsection{Is $N_{\rm GC}$ a Reliable Mass Diagnostic?}
\label{sec:discussion_Ngc}

The Disco Ball demonstrates that there are analogs to GCs in terms of luminosity and compactness that are extremely young in at least some UDGs. As these clusters age, they fade, affecting both the determination of the luminosity function and the number of GCs. This finding opens the question of the age distribution of clusters identified in other UDGs. Although state-of-the-art studies apply color criteria to help remove background, and perhaps also very young clusters, the criteria are generally loose and do not exclude intermediate age clusters that can still fade by $\sim$ 1 mag. For example, \cite{Saifollahi+2022} use a criterion of $0.4 < {\rm F475W} - {\rm F814W} < 1.5$, which according to our models only rejects clusters with ages $\lesssim 1$Gyr, and \cite{Shen_2021} use $0.31 < {\rm F606W} - {\rm F814W} < 0.484$, which only rejects clusters with ages $\lesssim 2$Gyr. Only in certain cases \citep[e.g.,][]{vdk18}, where spectroscopy is available and stellar population modeling can be done, have old ages been confirmed.

The inclusion of younger and more luminous clusters can affect various inferences. For example, the additional ``artificial" shift between the peak of the observed LF and the canonical GCLF of $\sim 1.0$ mag would cause a 37\% underestimation of the distance when using the GCLF turnover as a distance indicator. Regarding the estimate of $N_{\rm GC}$, we have already discussed the difference among the scenarios where the GCLF peak appears shifted due to incompleteness vs. a non-standard GCLF vs. where the peak is shifted due to fading. The difference in $N_{\rm GC}$ between the scenarios for the Disco Ball is about a factor of four. Some of the observed variations in $N_{\rm GC}$, such as those identified by \cite{forbes23}, could arise from age differences between GC populations. There may be true physical reasons for variations as well \citep{forbes25}, but the color cuts generally used to select GCs are insufficiently precise to fully limit the samples to ancient clusters and eliminate fading of intermediate age clusters as a possibility.
Evidence for unusual star formation histories in at least some UDGs is becoming more prevalent \citep{ferre,loraine}.

The presence of an NSC may also distort estimates of the galaxy's total mass derived from $N_{\rm GC}$. One widely accepted NSC formation scenario involves the amalgamation of infalling GCs into the galactic center. In this scenario, some GCs have already merged with the NSC, implying that the original number of GCs was higher than currently observed. So far, there is no statistical evidence of this effect when comparing the GC numbers in nucleated vs. non-nucleated UDGs \citep{marleau}.

\subsection{Are Some Red-Sequence UDGs Rotationally Supported?}

The Disco Ball provides a clear example of a large UDG that lies within the red sequence in the color-magnitude diagram that is rotationally supported. In contrast, other red-sequence UDGs are definitively not rotationally supported. 
DF44 in the Coma cluster, one of the best-studied UDGs in terms of its kinematic profile, does not exhibit bulk rotation \citep{vdk19}. 
One possibility is
that the large range in $v_{\rm rot}/\sigma$  among UDGs is related to the environment. The Disco Ball is located in a sparse environment  \citep[$D_{\rm 10} \sim 3$Mpc;][]{khim+24_nsc} while DF44 is in the Coma cluster. 
Our upcoming full analysis of the KCWI data, which focused observations on galaxies in sparser environments, may help address this possibility.

\subsection{Future Work}

To deepen our understanding of the Disco Ball, several avenues for future investigation are clear. First, obtaining higher S/N observations of the NSC will enable higher precision measurements of emission line ratios. Evidence for an IMBH in the NSC of the Disco Ball is currently marginal at best, but does prompt a question of the possible role of AGN activity in UDG evolution. Combined with future H$\alpha$ observations, these data will allow diagnostic assessments of massive central black hole activity within the system \citep[e.g.,][]{BPT}.
Second,
high-resolution imaging of emission-line clusters (or star-forming knots) would help us determine if these knots are genuinely bound stellar clusters or looser, transient star-forming regions. Similarly, further examination of the non-emission-line stellar clumps would allow us to determine whether these are truly analogs of systems identified in other UDGs. Third,  unresolved H{\small I} observations would provide a total gas mass measurement and place constraints on the role of gas and feedback in fueling current and future star formation, while spatially resolved observations might help answer why the emission line clusters are roughly aligned along the minor axis. One possibility is that there are weak, large pitch angle spiral arms in the Disco Ball and these may be visible in the gas distribution.
Finally, extending the spatially-resolved spectroscopic observations beyond the current field of view would allow us to use any additional emission line cluster to extend the rotation curve and provide a more direct measure of the full cluster population.

\section{Summary}
\label{sec:summary}

We present an in-depth study of the ``Disco Ball" (SMDG0038365-064207) --- a red sequence, rotationally-supported ultra-diffuse galaxy hosting a nuclear star cluster (NSC), multiple star clusters, and active star formation regions. Using spatially resolved spectroscopy obtained with the Keck Cosmic Web Imager (KCWI) on Keck II, we analyze its cluster population and kinematics and estimate its dynamical mass.  Our key results are:

\medskip
\noindent
$\bullet$ To a projected radius of 0.6$r_{\rm e}$, we identify 17 compact luminous clumps, including the NSC. The spectra of seven of these, including the NSC, show emission lines. The objects whose spectra do not exhibit emission lines we consider to be stellar clusters.
Their luminosity function peaks about one magnitude brighter than the canonical globular cluster luminosity function. We discuss a variety of possible reasons for this offset and, given the presence of recent star formation activity in the galaxy, favor a scenario where these are intermediate age clusters and therefore somewhat brighter than their $\sim$ 10Gyr old counterparts in other galaxies.

\medskip
\noindent
$\bullet$ 
Taking into account only radial incompleteness, we estimate
that the number of clusters ($N_{\rm GC}$) in the Disco Ball is
34 $\pm$ 11. The galaxy's halo mass ($M_{\rm h}$), estimated using the scaling relation between $N_{\rm GC}$ and $M_{\rm h}$, is $10^{11.25 \pm 0.26}$\Msol, within the range ($10^{11.01}$ to $10^{11.58}$\Msol)  of analogous estimates for the prototypical massive UDG, DF44 \citep{Saifollahi+2022,gannon24}.

\medskip
\noindent
$\bullet$ We find that the galaxy is rotationally supported. This is the first large, red sequence UDG where rotation is evident in the diffuse stellar population. Using the best-fit linear rotation curve, we measure a projected rotation velocity of 35.0 $\pm$ 6.7km s$^{-1}$ at 0.5$r_{\rm e}$, inclination corrected to 53.0 $\pm$ 10.2km sec$^{-1}$. Assuming no further increase in the rotation velocity beyond 0.5$r_{\rm e}$, we estimate that the dynamical mass within $r_{\rm e}$ is $10^{9.3 \pm 0.2}$ M$_\odot$ from the rotation formula (Equation \ref{eq:rotation_formula}) and $10^{9.6 \pm 0.2}$\Msol from the Wolf mass estimator. Extrapolating this value to obtain $M_{\rm h}$ adopting an NFW halo model, using the methodology of \cite{2023MNRAS.519..871Z}, and taking two different approaches, we calculate $M_{\rm h} = 10^{10.3\pm0.6}$ and  $10^{10.9\pm 0.6}$\Msol.  Combining these estimates with that obtained using $N_{\rm GC}$, we conclude that $M_{\rm h} = 10^{11.1\pm0.2}$\Msol.

\medskip
\noindent
$\bullet$ Assuming that the Disco Ball lies on the Baryonic Tully-Fisher Relation of \cite{Trachternach+09}, we find that it has baryonic mass of $10^{8.6 \pm 0.3}$\Msol. This value is nearly identical to our photometric estimate of the stellar mass ($10^{8.5\pm0.03}$\Msol), indicating that the galaxy is not a gas-dominated UDG, despite signs of recent or ongoing star formation. Future H{\small I} observations will determine if this inference is valid.

\medskip
\noindent
$\bullet$ The Disco Ball hosts a bright NSC, equivalent in stellar mass to the sum of $\sim$ 18 ``standard" globular clusters. We find a suggestion of broad [O III] emission lines, perhaps indicative of unusual kinematics, and an [O III] to [O II] line ratio consistent with shock or AGN ionization. Neither observation is statistically highly significant, but together they are least tempt us with the possibility of the presence of a weakly active nuclear source. 


\medskip
\noindent
$\bullet$ 
More generally, our observations of the Disco Ball bring into question two common assumptions in  the UDG literature. First, we conclude, although this is not the only viable interpretation,  that some of the stellar clusters we identify in the Disco Ball are of intermediate age and therefore more luminous than the globular clusters used in constructing the standard globular cluster luminosity function. Neglecting this age difference, if indeed it exists, will cause one to overestimate $N_{\rm GC}$ and force one to account for ``overluminous" clusters in some other way.  While this possibility may not be the explanation of overluminous clusters in every case where such clusters are found, we do caution that standard color selection criteria are insufficient to exclude the possibility. 

Second, because we find that the Disco Ball is rotationally supported, it demonstrates by example that even relatively quiescent UDGs can be rotationally rather than pressure supported. As such, the measured kinematics can be viewing-angle dependent. A face-on disk will present as a low velocity dispersion system, with a derived mass that is anomalously low relative to comparable systems.

\medskip
\noindent
The Disco Ball is a physically-large ultra-diffuse galaxy on the blue edge of the red sequence. Among the SMUDGes survey galaxies it does not present as unusual. Its size and mass place it among the more massive UDGs, but again in no unusual manner. Nevertheless, upon more detailed study we find that it holds a number of interesting surprises --- pointing to the possibility of even more surprises among this galaxy population once deep, spatially resolved spectroscopy becomes available for a larger set of these intriguing galaxies.

\begin{acknowledgments}
The authors acknowledge financial support from NSF AST-1713841 and AST-2006785. An allocation of computer time from the UA Research Computing High Performance Computing (HPC) at the University of Arizona and the prompt assistance of the associated computer support group is gratefully acknowledged.
This research utilizes images from the Dark Energy Camera Legacy Survey (DECaLS; Proposal ID 2014B-0404; PIs: David Schlegel and Arjun Dey). Full acknowledgment at https://www.legacysurvey.org/acknowledgment/.

\end{acknowledgments}

\software{
\texttt{Astropy }             \citep{astropy1, astropy2},
\texttt{dustmaps}            \citep{green},
\texttt{GALFIT }              \citep{peng},
\texttt{Matplotlib }          \citep{matplotlib},
\texttt{NumPy }               \citep{numpy},
\texttt{pandas }              \citep{pandas},
\texttt{scikit-learn }         \citep{sklearn},
\texttt{SciPy}                \citep{scipy1, scipy2},
\texttt{SEP}                \citep{sep},
\texttt{Source Extractor}     \citep{bertin},
\texttt{PYPHOT}             \citep{zenodopyphot}
}

\bibliography{references.bib}{}
\bibliographystyle{aasjournal}

\end{document}